# Generation of hallow vector beam by high-order cylindrical vector beams

**Brijesh Kumar Mishra, and Brijesh Kumar Singh**
*Department of Physics, School of Physical Sciences, Central University of Rajasthan, Ajmer 305817, Rajasthan, India*
*brijeshsingh@curaj.ac.in*

------------------------------------------------------------

*We propose a method for generating hollow beams using higher-order cylindrical vector modes of the form R-$TEM_{pl}$, where the radial index p is varied from 1 to 3 while the azimuthal index is fixed at l = 1. It is found that this scheme performs identically under incident illumination with either radial or azimuthal polarization. For this purpose, we use a focusing lens in combination with a diffractive optical element formed by a computer-generated hologram containing multiple alternate opaque and transparent regions. Based on vector diffraction theory, our analysis shows that the multi-zone amplitude mask redistributes the beam energy, thereby leading to the formation of a hollow beam. The proposed method provides control over the beam width which maintains a uniform dark core size after focusing through the various NA lens across all the higher order modes. Further the width of high intensity ring can be tuned by varying the NA of the focusing lens. This study shows that the proposed method is well suited for trapping particles or atoms while avoiding exposure to high central intensity, enabling improved contrast and resolution, facilitating ring-shaped ablation or heating, guiding atoms through dark regions to minimize thermal effects, and supporting information encoding using orbital angular momentum and other advanced optical applications.*
***Keyword:** Hallow beam, topological charge, beam order etc*

------------------------------------------------------------

## Introduction

Cylindrical vector beams, such as radially and azimuthally polarized beams, have attracted considerable attention due to their unique properties arising from their cylindrically symmetric electric field distributions. This behaviour fundamentally differs from that of conventional scalar beams, whose electric field direction remains uniform across the entire beam. Specifically, scalar and vector beams exhibit singularities on the beam axis in phase and polarization, respectively, which lead to an intensity null along the axis. Optical hollow beams are light beams characterized by zero or very low intensity at the centre, surrounded by ring-shaped regions of higher intensity, unlike Gaussian beams that exhibit a bright central spot. In cross-section, hollow beams resemble a donut shape intensity profile. In contrast, optical bottle beams feature bottle-shaped regions of low or zero intensity enclosed by three-dimensional regions of higher intensity [1]. Optical hollow beams are typically time-independent and it has been reported that optical bottle beams can exhibit dynamic behaviour along the transverse direction [2].

Numerous theoretical studies have predicted the existence of various types of cylindrical vector beams [3–6], and experimental realizations of these beams have been reported by several researchers [7]. Optical hollow beams and optical bottle beams can be generated using various methods, including axicons [8–10], spatial light modulators [11–13], holographic phase plates [14,15], moiré techniques [16], speckle patterns [17], diffractive optical elements [18], or computer-generated holograms [19]. A higher-order transverse mode beam with radially polarization can also be generated using a cavity mirror that provides polarization selectivity along with ring-shaped reflectivity modulation to suppress oscillation of other transverse modes [20,21]. This reflectivity modulation is a promising approach for selecting a specific transverse mode, as each mode exhibits a distinct intensity pattern.

Because of their null intensity close to the beam axis, which lowers scattering forces and permits the trapping of particles whose refractive index is lower than that of the surrounding medium [22,23], hollow beams, such as doughnut-shaped beams, have drawn a lot of attention [24–27]. These beams allow optical trapping using a single light beam [11,16,17] and are particularly effective for the manipulation of micron-sized light-absorbing particles [11,15,28], offering stable and controlled confinement. Furthermore, the high-intensity ring of a hollow beam acts as a repulsive pipe wall for particles confined in the central dark region, while the axial component of the thermal photophoretic force can drive particles along the beam channel, enabling guided transport and manipulation [15].

In this work, we present a comprehensive method for hollow-beam generation based on higher-order cylindrical vector modes. Unlike conventional approaches, which are primarily limited to lower-order modes and do not address hollow-beam formation across higher-order modes, previous techniques lack systematic control over the dark-core size and shape, as well as over the overall beam width. Moreover, the influence of numerical aperture on hollow-beam characteristics in high order regimes has remained largely unexplored. Our method overcomes these limitations by enabling hollow-beam generation using higher-order CV modes while preserving a constant dark-core size and shape. This invariance is maintained consistently under variations of the numerical aperture (e.g., NA = 0.01, 0.02, and 0.03). In addition, the proposed approach provides independent control over the width of intensity ring by varying the NA of the focusing lens without compromising the hollow-core geometry.

## Theory

In theory, Richards and Wolf's approach can numerically approximate the focussing of a vector beam modified by binary elements [29]. Youngworth and Brown explain the radial and longitudinal components of electric fields in cylindrical dimensions as [30,31], assuming that the incident electric field of the vector beam at the pupil plane in cylindrical coordinates is provided by:

$$e_\rho^{(s)}(\rho_s, z_s) = A \int_0^\alpha \cos^{1/2}\sin(2\theta) t_0(\theta) J_1(k\rho_s \sin\theta) e^{ikz_s \cos\theta} d\theta \quad (1)$$

$$e_\rho^{(s)}(\rho_s, z_s) = 2iA \int_0^\alpha \cos^{1/2}\sin^2\theta\, t_0(\theta) J_0(k\rho_s \sin\theta) e^{ikz_s \cos\theta} dZ \quad (2)$$

here, we express the field components using cylindrical coordinates (ρ, φ, z), the power coefficient (A) in Eqs. (1) and (2) is adjusted to 1 to maintain the same optical power in the image space. The apodization function $t_0(\theta)$ defines the relative amplitude of the electric field at the pupil plane. The maximum focusing angle α, is determined by $sin^{-1}(NA/n)$, where $k$ is the wavenumber in the image space $(k = 2\pi/\lambda)$, NA is the numerical aperture, and $J_m$ stands for the first kind Bessel function of order m. The following form of $t_0(\theta)$ for the R-TEM$_{pl}$ modes was used for a single doughnut-shaped beam that was produced as a paraxial solution of Maxwell's vector wave equation is given by:

$$l_0(\theta) = \left(\frac{\beta_0^2 \sin(\theta)}{\sin^2(\alpha)}\right) exp\left(-\frac{\beta_0^2 \sin^2(\theta)}{\sin^2(\alpha)}\right) L_p^l\left(\frac{2\beta_0^2 \sin^2(\theta)}{\sin^2(\alpha)}\right) \quad (3)$$

here, $L_p^l$ denotes the generalized Laguerre polynomial, and $\beta_0$ is defined as the truncation parameter, given by the ratio of the aperture radius to the beam waist.

Figure 1 illustrates the schematic of the proposed optical configuration. A radially polarized beam first impinges on a amplitude mask and is then focused by a lens with varying numerical aperture from 0.01 to 0.03. The mask, also shown in Figure 1, consists of a centrally located opaque region with zero transmission, surrounded by three concentric annular zones with transparent and opaque region.

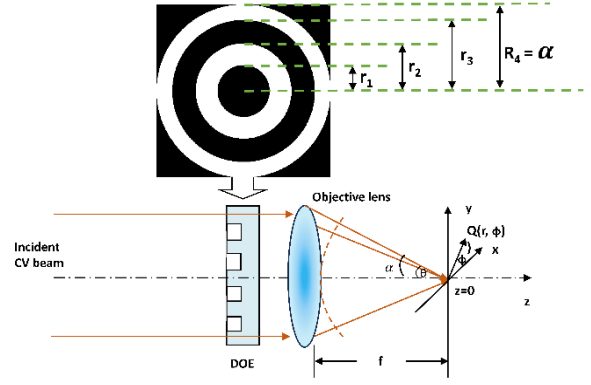

***Figure 1*** *Diagram of the focusing system, illustrating the propagation of the CV beam through the DOE and. the frontal layout of the amplitude DOE.*

In the schematic, the regions with 0 transmission are indicated in black, while those with 1 transmission are shown in white. The aperture half-angle corresponding to each annular zone, denoted by $\theta_i$ (i=1- 3) is given by $\theta_i = sin^{-1}(r_i NA)$, where $r_i$ represents the normalized radius of the $i^{th}$ zone. The influence of the amplitude mask on the focusing behaviour of the system is incorporated by modifying Eq. (1, 2) such that $t_0(\theta)$ is replaced with $t_0(\theta)$. $T(\theta)$ denotes the transmission function of the diffractive optical element. The explicit form of $\phi_0(\theta)$ is given by:

$$T(\theta) = \begin{cases} 0 & 0 \leq \theta < \theta_1, \quad \theta_2 \leq \theta < \theta_3, \\ 1 & \theta_1 \leq \theta < \theta_2, \quad \theta_3 \leq \theta < \alpha \end{cases} \quad (4)$$

To generate the hallow beam, the amplitude mask is optimized which enforces constructive interference in the high-intensity light barrier and destructive interference in the dark core region. To evaluate the beam-shaping performance of the proposed mask, numerical simulations are carried out based on Eqs. (4).

## Result

Figures 2-4 illustrate the effect of the amplitude mask on the normalized intensity profiles of high-order cylindrical vector beams corresponding to the R-TEM$_{11}$, R-TEM$_{21}$, and R-TEM$_{31}$ modes. These figures provide a comprehensive visualization of hollow-beam formation resulting from optical energy redistribution induced by amplitude mask. Each figure is organized in a consistent four-column format to facilitate direct comparison of the beam-focusing characteristics before and after modulation at different numerical aperture (NA) values. The first column depicts the unmodulated beam profile, serving as a reference. The second, third, and fourth columns present the modulated beam profiles for NA = 0.20, 0.25, and 0.35, respectively, arranged

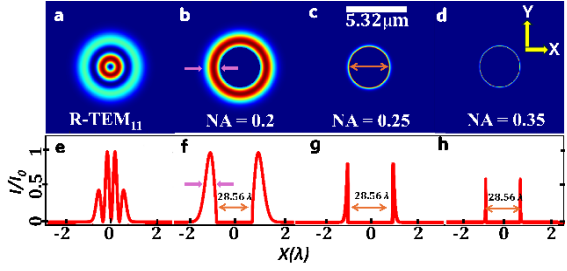

*Figure 2 simulation result for normalized intensity distributions of amplitude-modulated high-order cylindrical vector beams for the R-TEM$_{11}$ modes. Each figure follows a four-column layout: the first column shows the unmodulated beam profile, while the second, third, and fourth columns correspond to the modulated profiles for NA = 0.20, 0.25, and 0.35, respectively. The first row presents the 2D (a, b, c, d) transverse intensity distributions, and the second row shows the corresponding 1D (e, f, g, h) normalized total field intensity profiles.*

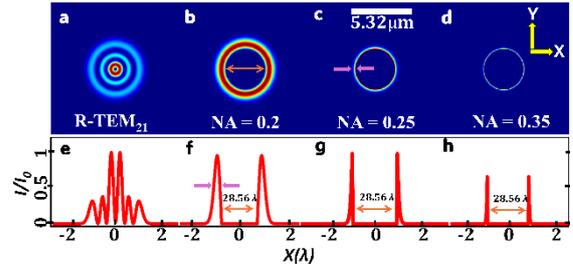

*Figure 3 simulation result for normalized intensity distributions of amplitude-modulated high-order cylindrical beams for the R-TEM$_{21}$ modes. Each figure follows a four-column layout: the first column shows the unmodulated beam profile, while the second, third, and fourth columns correspond to the modulated profiles for NA = 0.20, 0.25, and 0.35, respectively. The first row presents the 2D (a, b, c, d) transverse intensity distributions, and the second row shows the corresponding 1D (e, f, g, h) normalized total field intensity profiles.*

from left to right. The first row in each figure shows the two-dimensional (2D) transverse intensity distributions in the $xy$ plane at the focal plane (), highlighting the evolution of the hollow intensity structure with increasing NA. The second row presents the corresponding one-dimensional (1D) normalized total field intensity profiles along the transverse direction, plotted in red, which quantitatively demonstrate the redistribution of optical energy.

The influence of numerical aperture (NA) on the dark-core characteristics of high-order cylindrical vector beams was systematically investigated for the R-TEM$_{11}$, R-TEM$_{21}$, and R-TEM$_{31}$ modes under amplitude-modulated conditions. The analysis focuses on two key aspects: (i) the consistency of the dark core size in the modulated beam and (ii) the variation of the width of intensity ring as a function of NA of focusing system. For all considered modes and across the entire range of numerical apertures, the amplitude-modulated beam exhibits a remarkably stable dark core size. The transverse dark-core diameter remains invariant throughout the beam cross section and for all modes investigated, with a constant value of 28.56λ. This invariance confirms that the amplitude modulation effectively preserves the hollow-beam structure independent of the modal index and numerical aperture, demonstrating robust dark-core formation. However, the width of the intensity ring strongly dependent on the numerical aperture. A reduction in the intensity ring width is observed with increasing NA for all the modes.

Simulation results for the hallow beam formation of R-TEM$_{11}$ mode are shown in Figure 2. It is observed that with increasing NA from 0.2 to 0.35 the FWHM of the intensity ring decreases from 16.44 λ at NA = 0.20, 4.1 λ at NA = 0.25 and 1.74 λ at NA=0.35. A similar trend is observed for the R-TEM$_{21}$ mode (Figure 3), where the FWHM of the intensity ring width is reduced from 10.52 λ at NA = 0.20 to 4.26 λ at NA = 0.25 and further to 1.27 λ at NA = 0.35.

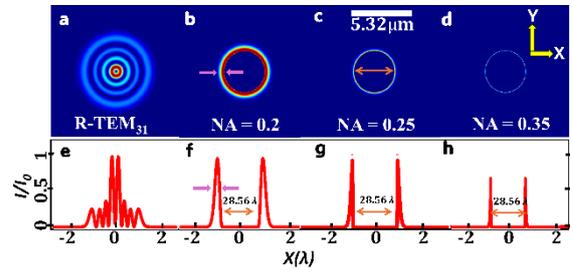

*Figure 4 simulation result for normalized intensity distributions of amplitude-modulated high-order cylindrical beams for the R-TEM$_{31}$ modes. Each figure follows a four-column layout: the first column shows the unmodulated beam profile, while the second, third, and fourth columns correspond to the modulated profiles for NA = 0.20, 0.25, and 0.35, respectively. The first row presents the 2D (a, b, c, d) transverse intensity distributions, and the second row shows the corresponding 1D (e, f, g, h) normalized total field intensity profiles.*

For the R-TEM$_{31}$ mode (Figure 4), the FWHM of the intensity ring decreases from 8.54 λ at NA = 0.20 to 2.50 λ at NA = 0.25 and reaches 1.56 λ at NA = 0.35.

Overall, these results clearly demonstrate that while the amplitude modulation ensures a mode-independent and NA-independent dark-core size, the FWHM of focused intensity ring can be precisely controlled by tuning the numerical aperture. This controllability, combined with the observed robustness of the dark core across different modes, highlights the potential of amplitude-modulated high-order cylindrical vector beams for applications requiring stable and tunable hollow intensity distributions.

## Conclusion

We have proposed a method for generating a hollow beam characterized by a dark central region of zero intensity, enclosed by a bright ring, using a high-order radially polarized beams focused with a low numerical aperture lens. Notably, the amplitude mask can be used regardless of the NA of the focusing lens, as the dark-core size of the resulting hollow beam remains unchanged.

Furthermore, the FWHM of intensity ring can be precisely tuned to localize at various positions with adjustable lengths and diameters. This proposed beam holds significant potential for applications in optical trapping and tweezers, cold atom manipulation, microscopy and imaging, optical communication, and laser material processing. These results clearly demonstrate the role of the amplitude mask in tailoring the spatial intensity distribution of high-order cylindrical vector beams and reveal the strong dependence of hollow-beam formation and focusing behaviour on the numerical aperture.